\documentclass[runningheads]{llncs}
\usepackage[linesnumbered,boxed,vlined]{algorithm2e}
\usepackage{graphicx}
\usepackage{amsmath}
\usepackage{amsfonts}
\usepackage{enumitem}
\usepackage{booktabs}
\usepackage{listings}
\usepackage{xcolor}

\definecolor{codegreen}{rgb}{0,0.6,0}
\definecolor{codegray}{rgb}{0.5,0.5,0.5}
\definecolor{codepurple}{rgb}{0.58,0,0.82}
\definecolor{backcolour}{rgb}{0.95,0.95,0.92}

\lstdefinestyle{mystyle}{
	backgroundcolor=\color{backcolour},
	commentstyle=\color{codegreen},
	keywordstyle=\color{magenta},
	numberstyle=\tiny\color{codegray},
	stringstyle=\color{codepurple},
	basicstyle=\ttfamily\footnotesize,
	breakatwhitespace=false,
	breaklines=true,
	captionpos=b,
	numbers=left,
	numbersep=5pt,
	showspaces=false,
	showstringspaces=false,
	showtabs=false,
	tabsize=2,
	prebreak=\mbox{{\color{codegray}\tiny$\searrow$}},
}

\begin{document}
	
	\title{Paralleling and Accelerating Arc Consistency Enforcement with Recurrent Tensor Computations}
	
	
	\titlerunning{Paralleling and Accelerating Arc Consistency Enforcement}
	%
	\author{Mingqi Yang\\
	National University of Singapore\\
	\email{mqyang@nus.edu.sg}}
	\authorrunning{Mingqi Yang}
	\institute{}

	\maketitle

	\begin{abstract}
		We propose a new arc consistency enforcement paradigm that transforms arc consistency enforcement into recurrent tensor operations. In each iteration of the recurrence, all involved processes can be fully parallelized with tensor operations. And the number of iterations is quite small. Based on these benefits, the resulting algorithm fully leverages the power of parallelization and GPU, and therefore is extremely efficient on large and densely connected constraint networks. 

	\end{abstract}

	\section{Introduction}
	In most existing architectures, arc consistency enforcement is designed as a sequential process on the constraint network, where the enforcement is first conducted on the current local parts of the network and then is propagated to the connected distant parts and so on~\cite{mackworth1977consistency,bessiere2001refining,lecoutre2007study,lecoutre2008enforcing,bessiere2005optimal,bessiere1994arc,vion2018mdd,wang2019arc}.
	This paradigm of arc consistency enforcement generally requires a $propagation$ queue and a $revision$ process, where the propagation queue maintains a list of variables or constraints to be revised and the revision fetches elements and removes the involved values that violate the arc consistency rule~\cite{rossi2006handbook}.

	This paradigm can be optimized to be much more incremental that remove many redundant operations between the two consecutive enforcements~\cite{balafoutis2008exploiting,boussemart2004revision,lecoutre2003exploiting,van2002ac,li2018new,mairy2014optimal}.
	But the $propagation$ manner is inherently sequential and can be difficult to parallelize.
	The drawback is obvious. On large or densely connected constraint networks, the chain of propagation conducted on the entire network can be quite long and time-consuming.

	To overcome this issue, we propose a new paradigm that transforms the traditional sequential propagation-based arc consistency enforcement into a group of parallelized tensor operations.
	We first reformalize the arc consistency enforcement as a recurrent process and then prove how the results of this recurrent process are equivalent to traditional arc consistency enforcement.
	The benefit of this new paradigm is that in each iteration of the recurrent process, the involved operations are independent and can be fully parallelized with tensor operations. And the number of recurrent steps is dramatically less in comparison to the number of propagation steps in the traditional paradigm.
	As a result, the proposed algorithm based on this paradigm can fully leverage the power of parallel computing and can be extremely efficient on large or densely connected constraint networks.

	\section{Notation}
	$dom(x)$ denotes the domain of the variable x.
	The tuple $(x, a)$ denotes the assignment $x=a$.
	Then the domain of all variables in a CSP can be represented as $D=\{(x, a)|x\in Vars, a\in dom(x)\}$, where $Vars$ is the set of all variables.
	$c_{xy}|_{(x, a)}$ denotes the set of all supports of $(x, a)$ on the constraint $c_{xy}$, i.e. $c_{xy}|_{(x, a)}=\{\tau[y]|\tau\in rel(c_{xy})\wedge \tau[x]=a\}$.
	$C_x$ denotes the set of all constraints involving the variable $x$.

	\section{Recurrent Arc Consistency (RAC) Enforcement}
	Given a CSP with the domain of all variables $D$ as defined previously, according to the definition of arc consistency, for any $D_{\bar{ac}}\subseteq D$,
	\begin{equation}
		\nonumber
		\begin{aligned}
			&\textrm{$D_{\bar{ac}}$ is arc consistent}\\
			\Leftrightarrow
			&\textrm{$\forall(x, a)\in D_{\bar{ac}}$($(x, a)$ is arc consistent)}\\
			\Leftrightarrow
			&\textrm{$\forall(x, a)\in D_{\bar{ac}}\forall c_{xy}\in C_x$($(x, a)$ has valid supports on $c_{xy}$)}\\
			\Leftrightarrow
			&\forall(x, a)\in D_{\bar{ac}}\forall c_{xy}\in C_x (c_{xy}|_{(x, a)}\cap D_{\bar{ac}}\neq\emptyset).
		\end{aligned}
	\end{equation}
	Generally, there can be more than one $D_{\bar{ac}}\subseteq D$ that is arc consistent for a CSP.
	Let $D_{\bar{AC}}$ be the set of all $D_{\bar{ac}}$ that is arc consistent and $D_{ac}=\bigcup_{D_{\bar{ac}}\in D_{\bar{AC}}}D_{\bar{ac}}$. Then, $D_{ac}\in D_{\bar{AC}}$, and $D_{ac}$ is the result of the arc consistency enforcement.
	There are many algorithms proposed to compute $D_{ac}$.
	In contrast with all these algorithms, we reformulate the computation of $D_{ac}$ (arc consistency enforcement) with completely recurrent tensor operations, which can fully leverage the power of parallel computations.



	Let $D_{\widetilde{ac}}=D\backslash D_{ac}$. According to the definition of arc consistency, we have
	\begin{lemma}
		\label{lemma:1}
		For any $(x,a)\in D$, if there exists $c_{xy}\in C_x$ and $D_{\widetilde{ac}}^{\prime}\subseteq D_{\widetilde{ac}}$ such that $c_{xy}|_{(x, a)}\subseteq D_{\widetilde{ac}}^{\prime}$, then $(x,a)\in D_{\widetilde{ac}}$.
	\end{lemma}

	We prove Lemma \ref{lemma:1} in Appendix \ref{proof:lemma:1}.
	Let $D_{\widetilde{ac}}^{\prime\prime}=D_{\widetilde{ac}}^{\prime}\cup\{(x,a)\}$.
	Then $D_{\widetilde{ac}}^{\prime\prime}\subseteq D_{\widetilde{ac}}$ as $D_{\widetilde{ac}}^{\prime}\subseteq D_{\widetilde{ac}}$ and $(x,a)\in D_{\widetilde{ac}}$.
	By setting $D_{\widetilde{ac}}^{\prime}=D_{\widetilde{ac}}^{\prime\prime}$, we can iteratively enlarge the set $D_{\widetilde{ac}}^{\prime}\subseteq D_{\widetilde{ac}}$.
	Based on this idea, we design the recurrent way of collecting $(x, a)$ with no valid support as follows.
	\begin{equation}
		\label{equ:iter_ac}
		\begin{aligned}
			\left\{
			\begin{array}{ll}
				D_{\widetilde{ac}}^{(0)}=\emptyset, k=0 \\
				D_{\widetilde{ac}}^{(k)}=D_{\widetilde{ac}}^{(k-1)}\cup\{(x,a)|\exists y, c_{xy}|_{(x, a)}\subseteq D_{\widetilde{ac}}^{(k-1)}\}, k\in\mathbb Z_+.
			\end{array}
			\right.
		\end{aligned}
	\end{equation}
The collection of $\{(x,a)|\exists y, c_{xy}|_{(x, a)}\subseteq D_{\widetilde{ac}}^{(k-1)}\}$ in each iteration can be potentially parallelized.
Next, we show that the recurrence in Equation \ref{equ:iter_ac} will always end with $D_{\widetilde{ac}}^{(K)}=D_{\widetilde{ac}}$.

	\begin{proposition}
		\label{prop:1}
		According to the computation of $D_{\widetilde{ac}}^{(k)}$ in Eq.\ref{equ:iter_ac},
		\begin{enumerate}[noitemsep]
			\item
			$\forall k\in\mathbb Z_+(D_{\widetilde{ac}}^{(k)}\subseteq D_{\widetilde{ac}})$;
			\item
			there always exists $K$ such that,
			\begin{enumerate}
				\item
				$\{(x,a)|\exists y, c_{xy}|_{(x, a)}\subseteq D_{\widetilde{ac}}^{(K)}\}\backslash D_{\widetilde{ac}}^{(K)}=\emptyset$;
				\item
				$D_{\widetilde{ac}}^{(K)}=D_{\widetilde{ac}}$.
			\end{enumerate}
		\end{enumerate}
	\end{proposition}

	We prove Proposition \ref{prop:1} in Appendix \ref{proof:prop:1}.
	Therefore, Eq.\ref{equ:iter_ac} acts as an iterative manner to enforce arc consistency.
	When $D_{\widetilde{ac}}^{(K)}=D_{\widetilde{ac}}^{(K+1)}$, the iteration ends with $D_{ac}=D\backslash D_{\widetilde{ac}}^{(K)}$.
	More interestingly, Eq.\ref{equ:iter_ac} indicates the availability of paralleling the arc consistency enforcement as we can collect the element in $\{(x,a)|\exists y, c_{xy}|_{(x, a)}\subseteq D_{\widetilde{ac}}^{(k-1)}\}$ in each iteration simultaneously. We make the best use of parallelization by designing tensor operations.

	Before that, there is an interesting insight that indicates a possible optimization of increment among iterations of the recurrent process.

	\begin{proposition}
		\label{prop:2}
		Let $V_{\widetilde{ac}}^{(k)}=D_{\widetilde{ac}}^{(k)}\backslash D_{\widetilde{ac}}^{(k-1)}$, then
		\begin{enumerate}[noitemsep]
			\item
			$\forall(x,a)\in V_{\widetilde{ac}}^{(k)}\forall c_{xy}\in C_x(c_{xy}|_{(x, a)}\backslash D_{\widetilde{ac}}^{(k-2)}\neq\emptyset)$;
			\item
			$\forall(x,a)\in V_{\widetilde{ac}}^{(k)}\exists c_{xy}\in C_x(c_{xy}|_{(x, a)}\backslash D_{\widetilde{ac}}^{(k-2)}\subseteq V_{\widetilde{ac}}^{(k-1)})$.
		\end{enumerate}
	\end{proposition}

	We prove Proposition \ref{prop:2} in Appendix \ref{proof:prop:2}.
	Proposition \ref{prop:2} shows that all $(x, a)$ to be removed in the current iteration are caused by the loss of the supports removed in the previous iteration.
	So, we can make the whole enforcement process as in Equation \ref{equ:iter_ac} incremental by maintaining the set of variables with a changed domain and then only checking and removing inconsistent $(x, a)$ by only testing variables in this set.


	\begin{figure*}[h]
		\centering
		\includegraphics[width=0.45\textwidth]{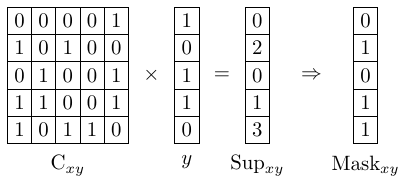}
		\caption{A variable $y$ is represented as a 1d array indexed by the values in $dom(y)$. In this array, $y[a]=1$ represents $y$ has the value $a$, and $y[a]=0$ represents not. Similarly, a constraint $C_{xy}$ is represented as a 2d array. $Sup_{xy}[a]$ represents the number of collected supports of $(x, a)$ on the constraint $C_{xy}$.}
		\label{fig:ac_single}
	\end{figure*}
	
	\section{RAC Enforcement with Tensor Accelerating (RTAC)}
	\begin{figure*}[t]
		\centering
		\includegraphics[width=1.0\textwidth]{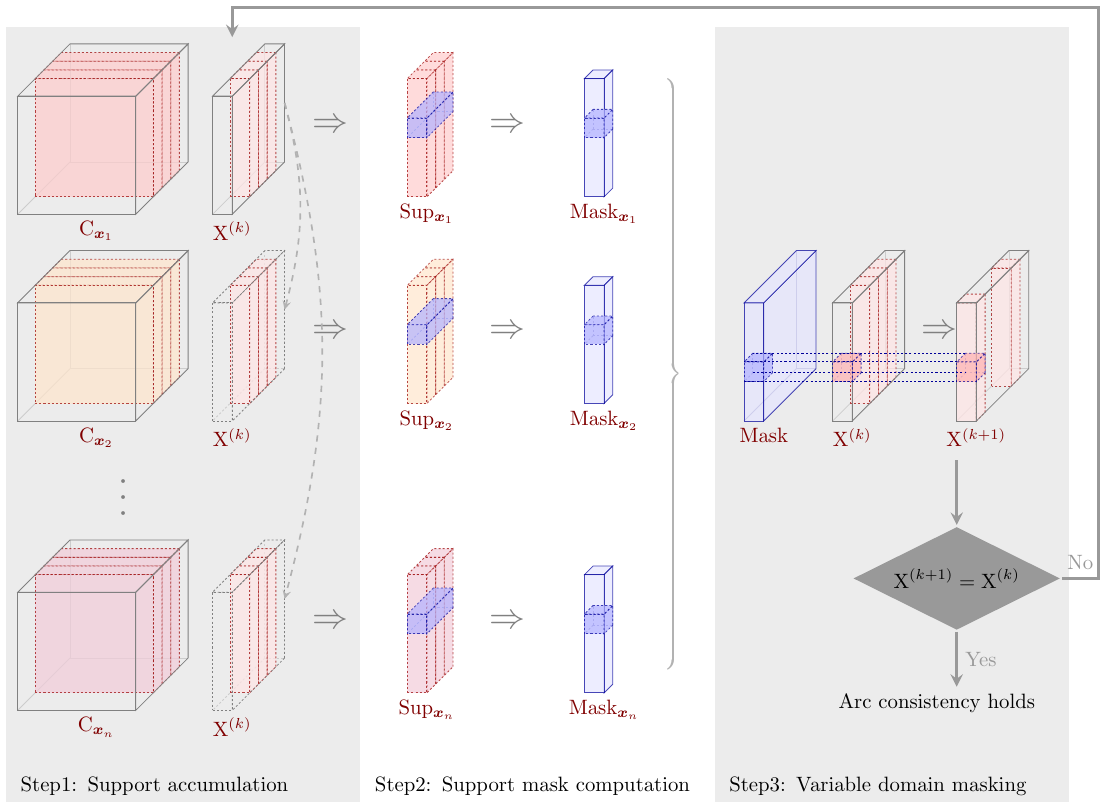}
		\caption{The illustration of arc consistency enforcement with tensor parallzation.}
		\label{fig:pipline}
	\end{figure*}
	In this section, we design our algorithm based on the recurrent arc consistency paradigm and accelerating each iteration, i.e. collecting $\{(x,a)|\exists y, c_{xy}|_{(x, a)}\subseteq D_{\widetilde{ac}}^{(k-1)}\}$, with tensor parallelization.

	We first introduce the basic idea of support collection with tensor computation with the simplest single constraint case as in Figure \ref{fig:ac_single}.
	Then, the whole support collecting on all constraints and variables is achieved by stacking multi-dimensional tensors and can be computed simultaneously as Step 1 in Figure \ref{fig:pipline}.
	The whole recurrent tensor arc consistency (RTAC) enforcement pipeline is as Figure \ref{fig:pipline}, and the corresponding pseudocode is as Algorithm \ref{alg:ac}.
	Algorithm \ref{alg:ac} only relies on a small group of basic tensor operations as follows where we use T to represent a given tensor:

	\begin{itemize}[noitemsep]
		\item T.\textbf{sum}($dim$):
		Returns the sum of each row of T in the given dimension $dim$.
		\item T.\textbf{any}():
		Tests if any element in T evaluates to True.
		\item T.\textbf{nonzero}():
		Return the indices of all non-zero elements of T.
		\item T.\textbf{dim\_expand}($dim$):
		Returns a tensor with a dimension of size one inserted to T at the specified $dim$.
		\item T.\textbf{dim\_reduct}($dim$):
		Returns a tensor with the specified $dim$ of size one of T removed.
		\item \textbf{where}($condition$, $x$, $y$):
		Return a tensor of elements selected from either $x$ or $y$, depending on $condition$.
	\end{itemize}

These operations are all well provided in most tensor computing/deep learning frameworks e.g., Pytorch, Jax, TensorFlow, Numpy, etc.
For ease of implementation, we build our program upon the popular deep learning framework PyTorch with the sacrifice of some computation efficiency.
The actual amount of code is quite little thanks to the well-developed tensor computing frameworks.
We give all codes in the Appendix \ref{app:code}.

	\SetKwComment{Comment}{/* }{ */}

	\begin{algorithm}
		\caption{Arc consistency enforcement}\label{alg:ac}
		\SetKwFunction{FtensorAC}{tensorAC}
		\SetKwFunction{Ftrevise}{tensorRevise}
		\SetKwProg{Fn}{Def}{:}{}
		\Fn{\FtensorAC{Vars, @changed}}{
			\#Vals$^{pre}$ = Vars.\textbf{sum}(1)\;
			\While{$|\textrm{@changed}| \neq 0$}{
				Vars=\Ftrevise{Vars, @changed}\;
				\#Vals = Vars.\textbf{sum}(1)\;
				\If{(\#Vals $==$ zero$^n$).\textbf{any}()}{
					\textbf{throw inconsistency}\;
				}
				@changed = (\#Vals $\neq$ \#Vals$^{pre}$).\textbf{nonzero}()\;
				\#Vals$^{pre}$ = \#Vals\;
			}
			\KwRet Vars\;
		}

		\SetKwProg{Fn}{Def}{:}{}
		\Fn{\Ftrevise{Vars, @changed}}{
			Cons = Cons[*, @changed, *, *]\;
			Vars = Vars[@changed, *].\textbf{dim\_expand}(2)\;
			supp = (Cons $\times$ Vars).\textbf{dim\_reduct}(-1)\;
			supp = \textbf{where}(supp $>$ 1, one$^{nnd}$[*, 0: $|$@changed$|$, *], supp)\;
			Vars = \textbf{where}(supp.\textbf{sum}(1) $\neq$ $|$@changed$|$, zero$^{nd}$, Vars)\;
			\KwRet Vars\;
		}
	\end{algorithm}

\section{Experiments}
\subsection{Configuration}
To ensure a fair comparison as much as possible, we implement the state-of-the-art sequential arc consistency enforcement algorithm AC3 with Python + JIT since the pure Python program is slow.
Then, we implement RTAC with Python + PyTorch.
For the hardware part, we use CPU: I9-10900K and GPU: RTX3090.

\subsection{Benchmark}
We use the randomly generated binary constraint satisfaction problems since they can easily be generated with different scales, i.e. different number of variables, constraints, etc, to be used to conduct our ablation studies.
The constraint network topology is generated randomly with manually setting constraint density.
Specifically, for a number of $n$ variables and a given constraint density $d$.
There will be $\frac{n\times(n-1)}{2}$ pair of variables, and each pair of them is assigned with a constraint with the possibility of $d$.
We generate a total of 25 random CSPs with the number of variables $\{100, 250, 500, 750, 1000\}$ and the densities $\{0.1, 0.25, 0.5, 0.75, 1.0\}$.

\begin{figure*}[h]
	\centering
	\includegraphics[width=1.0\textwidth]{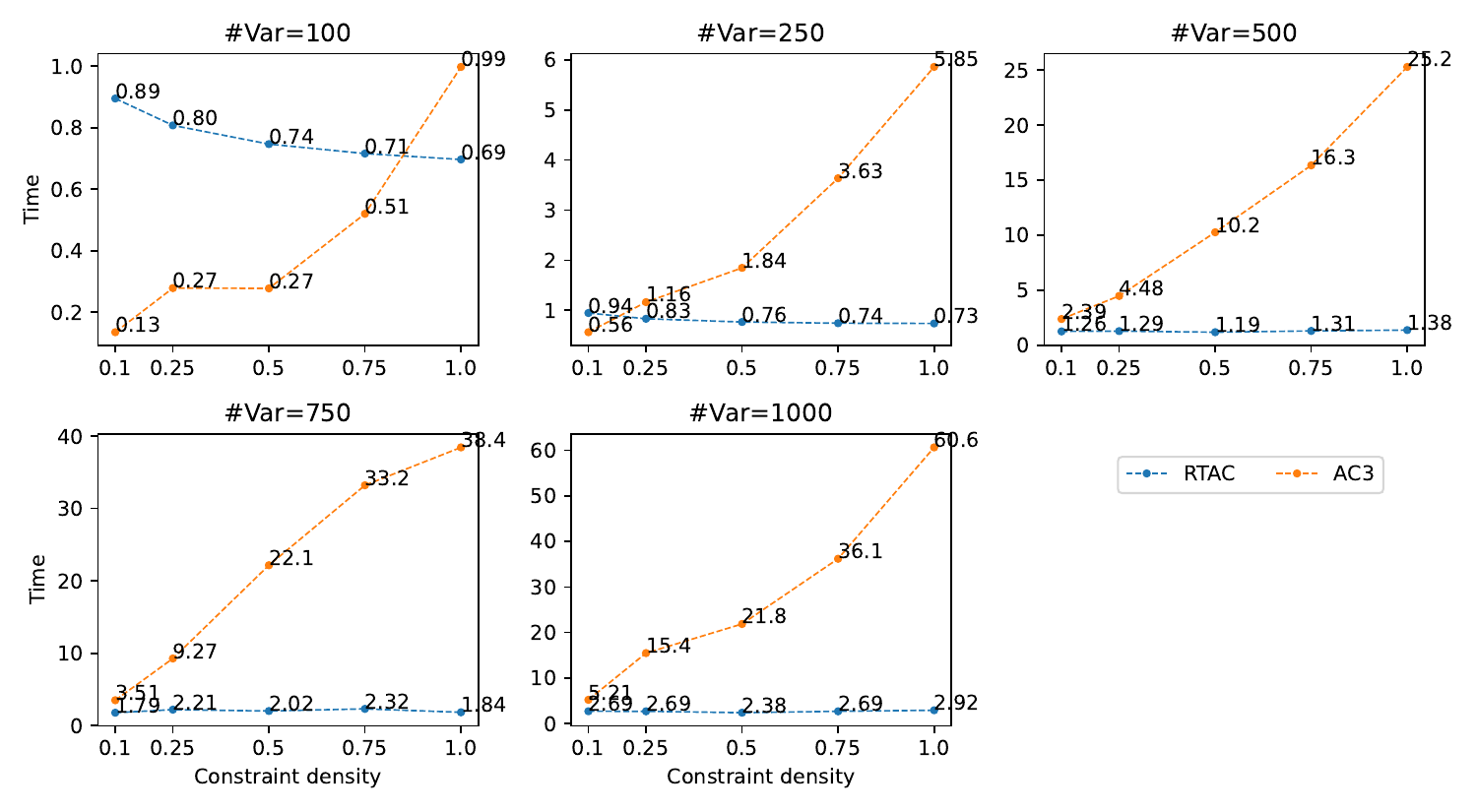}
	\caption{Running time (ms) of one assignment in backtrack search. The results are an average of 50K assignments.}
	\label{fig:ablation_plot}
\end{figure*}

\subsection{Result}
Comparing the efficiency of our proposed RTAC and the traditional sequential algorithms can be tricky for the following reasons: First, the arc consistency enforcements of two algorithms do not run on the same device. The former runs on GPU and the latter runs on CPU. Thus their efficiency is partially decided by the performance of the hardware. Second, although RTAC can be built from scratch to ensure the best efficiency at every detailed implementation, we built it upon the deep learning framework for ease of implementation. Therefore its performance is mostly decided by the efficiency of the selected framework.

As noticed in the above concerns, the comparisons between RTAC and AC3 can be less rigorous.
But the empirical results in Fig.\ref{fig:ablation_plot} and Tab.\ref{tab:ablation_table} indicate the following two guarantees:
First, the increase of time-consuming of RTAC is even unnoticeable when increasing the number of variables or the density of constraints; Second, RTAC has great potential on large and densely connected CSPs.

\begin{table}[h]
	\centering
	\caption{A statistics of the number of revisions (denoted by \#Revision) in AC3 and the number of recurrences (denoted by \#Recurrence) in RTAC. The results are an average of 50K assignments.}
	\label{tab:ablation_table}
	\begin{tabular}{cc|cc}
		\toprule
		\#Variable & Density & \#Revision   & \#Recurrence \\
		\midrule
		100  & 0.10        & 307.6     & 4.509        \\
		100  & 0.25        & 626.4     & 4.103        \\
		100  & 0.50        & 965.2     & 3.752        \\
		100  & 0.75        & 1612.4    & 3.573        \\
		100  & 1.00        & 2714.1    & 3.462        \\
		250  & 0.10        & 1152.0    & 4.804        \\
		250  & 0.25        & 2532.6    & 4.167        \\
		250  & 0.50        & 4629.6    & 3.794        \\
		250  & 0.75        & 7881.9    & 3.617        \\
		250  & 1.00        & 12405.6   & 3.441        \\
		500  & 0.10        & 3250.9    & 4.620        \\
		500  & 0.25        & 7619.8    & 4.126        \\
		500  & 0.50        & 18793.8   & 3.952        \\
		500  & 0.75        & 28218.4   & 3.728        \\
		500  & 1.00        & 42557.7   & 3.455        \\
		750  & 0.10        & 6195.7    & 4.766        \\
		750  & 0.25        & 13768.6   & 4.020        \\
		750  & 0.50        & 36220.6   & 3.940        \\
		750  & 0.75        & 61171.7   & 3.703        \\
		750  & 1.00        & 71509.8   & 3.597        \\
		1000 & 0.10        & 8322.2    & 4.831        \\
		1000 & 0.25        & 24544.3   & 4.381        \\
		1000 & 0.50        & 39707.7   & 4.048        \\
		1000 & 0.75        & 65446.2   & 3.755        \\
		1000 & 1.00        & 107680.5  & 3.556       \\
		\bottomrule
	\end{tabular}
\end{table}

\section{Conclusion}
We propose a new arc consistency enforcement paradigm and theoretically prove the equivalence of its results with the definition of arc consistency.
The induced algorithm RTAC fully leverages the power of parallelization and GPU, showing its efficiency on large and densely connected constraint networks.

\bibliographystyle{abbrv}
\bibliography{reference}

	\newpage
	\appendix

	\section{Proof of Lemma \ref{lemma:1}}
	\label{proof:lemma:1}
	\begin{proof}
		$c_{xy}|_{(x, a)}\subseteq D_{\widetilde{ac}}^{\prime}\subseteq D_{\widetilde{ac}}$, So $c_{xy}|_{(x, a)}\cap D_{ac}=\emptyset$, which means $(x,a)$ has no valid support on $c_{xy}$, and therefore $(x,a)$ is not arc consistent. Hence, $(x,a)\notin D_{ac}$ and $(x,a)\in D_{\widetilde{ac}}$.
	\end{proof}

	\section{Proof of Proposition \ref{prop:1}}
	\label{proof:prop:1}
	\begin{proof}
		(1) When $k=0,D_{\widetilde{ac}}^{(0)}=\emptyset\subseteq D_{\widetilde{ac}}$.
		Suppose when $k=l$, $D_{\widetilde{ac}}^{(l)}\subseteq D_{\widetilde{ac}}$.
		According to Prop.\ref{lemma:1}, $\{(x,a)|\exists y, c_{xy}|_{(x, a)}\subseteq D_{\widetilde{ac}}^{(l)}\}\subseteq D_{\widetilde{ac}}$.
		Then, when $k=l+1$, $D_{\widetilde{ac}}^{(l+1)}=D_{\widetilde{ac}}^{(l)}\cup\{(x,a)|\exists y, c_{xy}|_{(x, a)}\subseteq D_{\widetilde{ac}}^{(l)}\}\subseteq D_{ac}$.
		So, $\forall k\in\mathbb Z_+, D_{\widetilde{ac}}^{(k)}\subseteq D_{\widetilde{ac}}$.

		(2.a) If there is no such $K$, in other words, for any $k\in\mathbb Z_+$, $\{(x,a)|\exists y, c_{xy}|_{(x, a)}\subseteq D_{\widetilde{ac}}^{(k)}\}\backslash D_{\widetilde{ac}}^{(k)}\neq\emptyset$, then $D_{\widetilde{ac}}^{(0)}\subset D_{\widetilde{ac}}^{(1)}\subset\dots\subset D_{\widetilde{ac}}^{(+\infty)}$. So $\lim_{k\rightarrow+\infty}|D_{\widetilde{ac}}^{(k)}|=+\infty$, which is inconsistent with the fact that $D_{\widetilde{ac}}^{(k)}\subseteq D_{\widetilde{ac}}$, thus reaching conflicts.

		(2.b) As $D_{\widetilde{ac}}^{(k)}\subseteq D_{\widetilde{ac}}$ for any $k$, we have
		\begin{equation}
			\nonumber
			\begin{aligned}
				&\{(x,a)|\exists y, c_{xy}|_{(x, a)}\subseteq D_{\widetilde{ac}}^{(K)}\}\backslash D_{\widetilde{ac}}^{(K)}=\emptyset\\
				\Leftrightarrow
				&\{(x,a)|\exists y, c_{xy}|_{(x, a)}\subseteq D_{\widetilde{ac}}^{(K)}\}\subseteq D_{\widetilde{ac}}^{(K)}\\
				\Leftrightarrow
				&\forall(x, a)(\exists c_{xy}\in C_x(c_{xy}|_{(x, a)}\subseteq D_{\widetilde{ac}}^{(K)})\rightarrow(x, a)\in D_{\widetilde{ac}}^{(K)})\\
				\Leftrightarrow
				&\forall(x, a)((x, a)\in (D\backslash D_{\widetilde{ac}}^{(K)})\rightarrow\forall c_{xy}\in C_x(c_{xy}|_{(x, a)}\cap (D\backslash D_{\widetilde{ac}}^{(K)})\neq\emptyset))\\
				\Leftrightarrow
				&\textrm{$\forall(x, a)\in D\backslash D_{\widetilde{ac}}^{(K)}\forall c_{xy}\in C_x$($(x, a)$ has valid supports on $c_{xy}$)}\\
				\Leftrightarrow
				&\textrm{$D\backslash D_{\widetilde{ac}}^{(K)}$ is arc consistent}\\
				\Rightarrow
				&D\backslash D_{\widetilde{ac}}^{(K)}\subseteq D_{ac}=D\backslash D_{\widetilde{ac}}\\
				\Leftrightarrow
				&D_{\widetilde{ac}}\subseteq D_{\widetilde{ac}}^{(K)}\\
				\Leftrightarrow
				&D_{\widetilde{ac}}=D_{\widetilde{ac}}^{(K)}
			\end{aligned}
		\end{equation}

	\end{proof}

	\section{Proof of Proposition \ref{prop:2}}
	\label{proof:prop:2}
	\begin{proof}
		(1) For any $(x,a)\in V_{\widetilde{ac}}^{(k)}$, if there exists $c_{xy}\in C_x$ with $c_{xy}|_{(x, a)}\backslash D_{\widetilde{ac}}^{(k-2)}=\emptyset$, aka, $\exists y, c_{xy}|_{(x, a)}\subseteq D_{\widetilde{ac}}^{(k-2)}$, then $(x,a)\in D_{\widetilde{ac}}^{(k-1)}$ according to Eq.\ref{equ:iter_ac}, thus reaching conflicts.

		(2)
		\begin{equation}
			\nonumber
			\begin{aligned}
				V_{\widetilde{ac}}^{(k)}
				&=D_{\widetilde{ac}}^{(k)}\backslash D_{\widetilde{ac}}^{(k-1)}\\
				&=D_{\widetilde{ac}}^{(k-1)}\cup\{(x,a)|\exists c_{xy}\in C_x, c_{xy}|_{(x, a)}\subseteq D_{\widetilde{ac}}^{(k-1)}\}\backslash D_{\widetilde{ac}}^{(k-1)}\\
				&\subseteq\{(x,a)|\exists c_{xy}\in C_x, c_{xy}|_{(x, a)}\subseteq D_{\widetilde{ac}}^{(k-1)}\}\\
				&=\{(x,a)|\exists c_{xy}\in C_x, c_{xy}|_{(x, a)}\subseteq D_{\widetilde{ac}}^{(k-2)}\cup V_{\widetilde{ac}}^{(k-1)}\}\\
				&=\{(x,a)|\exists c_{xy}\in C_x, c_{xy}|_{(x, a)}\backslash D_{\widetilde{ac}}^{(k-2)}\subseteq V_{\widetilde{ac}}^{(k-1)}\}
			\end{aligned}
		\end{equation}
		Hence, $\forall(x,a)\in V_{\widetilde{ac}}^{(k)}\exists c_{xy}\in C_x(c_{xy}|_{(x, a)}\backslash D_{\widetilde{ac}}^{(k-2)}\subseteq V_{\widetilde{ac}}^{(k-1)}$).
	\end{proof}

\section{Pseudocode of Backtrack search}

\begin{algorithm}
	\caption{Backtrack search}\label{alg:btsearch}
	\SetKwFunction{Fmain}{main}
	\SetKwFunction{Fdfs}{dfs}
	\SetKwFunction{Finit}{init}
	\SetKwFunction{Fassign}{assign}
	\SetKwFunction{Fheuristics}{heuristics}
	\SetKwProg{Fn}{Def}{:}{}
	\Fn{\Fmain{}}{
		\Finit{}\;
		\FtensorAC{Vars, $[0: |Vars|]$}\;
		\Fdfs{0, Vars}\;
	}
	\SetKwProg{Fn}{Def}{:}{}
	\Fn{\Fdfs{level, Vars}}{
		\If{$level==n$}{
			\textbf{find answer}\;
		}
		idx = \Fheuristics{}\;
		\For{$Val\in Var[idx].\textbf{nonzero}()$}{
			Vars = \Fassign{idx, Val, Vars}\;
			Vars = \FtensorAC{Vars, [idx]}\;
			\If{\Fdfs(level+1, Vars)}{
				\KwRet True\;
			}
		}
		\KwRet False\;
	}
	\SetKwProg{Fn}{Def}{:}{}
	\Fn{\Finit{}}{
		\textbf{Prepare} $\textrm{Cons}\in\{0, 1\}^{n\times n\times d\times d}$\;
		\textbf{Prepare} $\textrm{Vars}\in\{0, 1\}^{n\times d}$\;
		\textbf{Prepare} $\textrm{zero}^n\in 0^n$\;
		\textbf{Prepare} $\textrm{zero}^{nd}\in 0^{n\times d}$\;
		\textbf{Prepare} $\textrm{one}^{nnd}\in 1^{n\times n\times d}$\;
		\textbf{Prepare} $\textrm{I}^{nn}=$ Identity matrix\;
	}
	\SetKwProg{Fn}{Def}{:}{}
	\Fn{\Fassign{idx, Val, Vars}}{
		I$^{nn}$[idx][idx] = 0\;
		Vars = I$^{nn}$ $\times$ Vars\;
		I$^{nn}$[idx][idx] = 1\;
		Vars[idx][Val] = 1\;
		\KwRet Vars\;
	}
\end{algorithm}

\section{Source Code of RTAC}
\label{app:code}
The amount of source code of RTAC is dramatically small when implemented upon the deep learning libraries.
Below is the code of RTAC built upon PyTorch.

\begin{lstlisting}[language=python, caption=The code of RTAC implemented with PyTorch., captionpos=b, float=*t]
import torch

class ACEnforcer:
    def __init__(self, cons_map, n_vars, n_dom):
        self.cons_map = cons_map
        self.n_mask0 = torch.zeros(n_vars).to(device)
        self.nnd_mask1 = torch.ones((n_vars, n_vars, n_dom)).to(device)
        self.nd_mask0 = torch.zeros((n_vars, n_dom)).to(device)

    def ac_enforcer(self, vars_map, changed_idx):
        n_idx = changed_idx.shape[0]
        vars_map_pre = vars_map.sum(1)
        while n_idx != 0:
            nkd = torch.matmul(self.cons_map[:, changed_idx, :, :], vars_map[changed_idx, :].unsqueeze(2)).squeeze(-1)
            nd = torch.where(nkd > 1, self.nnd_mask1[:, : n_idx, :], nkd).sum(1)
            vars_map = torch.where(nd != n_idx, self.nd_mask0, vars_map)
            vars_map_sum = vars_map.sum(1)
            if (vars_map_sum == self.n_mask0).any():
                return None
            changed_idx = (vars_map_sum != vars_map_pre).nonzero(as_tuple=True)[0]
            n_idx = changed_idx.shape[0]
            vars_map_pre = vars_map_sum
        return vars_map
\end{lstlisting}
	
\end{document}